# THE TOPOLOGY OF RR LYRAE INSTABILITY STRIP AND THE OOSTERHOFF DICHOTOMY

Giuseppe Bono

Osservatorio Astronomico di Trieste, Via G.B. Tiepolo 11, 34131 Trieste, Italy;
bono@oat.ts.astro.it

Filippina Caputo

Istituto di Astrofisica Spaziale, Via E. Fermi 21, 00044 Frascati, Italy;
caputo@saturn.ias.fra.cnr.it

and

Marcella Marconi

Dipartimento di Fisica, Univ. di Pisa, Piazza Torricelli 2, 56100 Pisa, Italy;
marcella@astr1pi.difi.unipi.it

## ABSTRACT

Convective pulsating models with $Y = 0.24$, stellar mass $M = 0.65 M_\odot$, $0.75 M_\odot$, under selected assumptions about luminosities and effective temperatures, are used together with stellar atmosphere computations to get the predicted dependence of RR Lyrae instability strip on stellar mass in the observative plane $(M_V, B - V)$. The comparison with observed variables belonging to the globular clusters M3 and M15 is presented. It appears that in M3 (Oosterhoff type I) the transition between *ab*-type and *c*-type RR Lyrae takes place along the fundamental blue edge, whereas in M15 (Oosterhoff type II) it occurs near the first overtone red edge. This result supports the key role of the *hysteresis mechanism* originally proposed by van Albada & Baker for explaining the Oosterhoff dichotomy.

# 1. INTRODUCTION

More than half a century ago it was shown (Oosterhoff 1939, 1944) that on the ground of the average periods of *ab*-type RR Lyrae stars the Galactic globular clusters (GGC) can be subdivided into two groups: Oosterhoff type I (OoI) and Oosterhoff type II (OoII) clusters, with $<P_{ab}>$ clustered around $0.55d$ and $0.65d$, respectively. This early classification appeared very soon in connection with other observational parameters such as the relative number of *c*-type variables, the metallicity and the horizontal branch (HB) morphology of the parent cluster. The picture which emerged on the whole is that the GGC system suffers an indisputable dichotomy. Metal-intermediate OoI clusters show a predominance of RR*ab* over RR*c* stars and red or evenly populated HB, whereas metal-poor OoII clusters have almost the same number of RR*c* and RR*ab* and blue HB morphology.

Given such an observational evidence it results quite clearly that the solution of the Oosterhoff dichotomy lies in taking simultaneously into account both evolutionary and pulsational constraints derived from the computations of HB models and pulsating models respectively. Along this line Bono, Caputo, & Stellingwerf (1994) have recently discussed the features of Oosterhoff groups within the framework of both an up-to-date pulsational scenario (Bono & Stellingwerf 1994, hereinafter [BS]) and canonical synthetic horizontal branch computations (Caputo et al. 1993). They showed that the observed mean pulsational properties are in agreement with the prescription of a constant helium abundance ($Y \sim 0.24$) among GGC if in OoI clusters the RR*ab*/R*c* transition occurs near the fundamental blue edge, whereas in OoII clusters this transition takes place close to the first overtone red edge. According to such a result the different mean pulsational properties of cluster variables belonging to both Galactic and external galaxies globulars are caused by the same combination of HB evolutionary tracks with a hysteresis mechanism in the modal stability of pulsating stars..

However, the quoted investigation was based on BS models, as computed for $Y = 0.30$ and $M = 0.65 M_\odot$, adopting suitable but temptative evaluations of the dependence of the strip on both helium content and stellar mass. In order to bridge the traditional gap between pulsational and evolutionary parameters, in this paper we present and discuss two new sets of pulsating models calculated according to current evolutionary evaluations for HB stars, namely $M = 0.65 M_\odot$, $0.75 M_\odot$ and $Y = 0.24$ (Marconi 1994, 1995).

Since the main aim of the present investigation is to allow a comparison between the overall topology of the RR Lyrae instability strip and the observed distribution of cluster variables, both blue and red fundamental edges (FBE, FRE) as well as blue and red first overtone edges (HBE, HRE) have been transformed into the observative plane $(M_V, B - V)$, by using bolometric corrections (BC) and color-temperature relations (CTR) provided by Kurucz (1992).

On this basis we will discuss the prototypes of the two Oosterhoff groups − $M3$ (OoI) and M15 (OoII) − even though the basic quantities for the comparison with variables of any other globular cluster are provided.

## 2. THE CASES OF M3 AND M15

Figure 1 shows the theoretical instability strip as computed for $Y = 0.24$, $M = 0.65 M_\odot$ (dashed line), and $M = 0.75 M_\odot$ (solid line): from left to right, the various lines represent HBE, FBE, HRE, and FRE (see also Table 1). A detailed discussion of these new models will be given in a forthcoming paper (Bono & Marconi 1995), but the interested reader is referred to BS for what concerns the adopted numerical and physical inputs. Table 2 gives the edges for fundamental and first overtone modes transformed into the observative plane by using three different choices of metal content ($Z = 0.0001$, 0.0004, and 0.001). Although, the envelope models were computed by assuming a fixed metal content ($Z = 0.001$) the theoretical instability strip can be transformed by using different Z values since the pulsation properties from $Z = 0.0001$ to $Z = 0.001$ do not depend on metallicity. It is worth mentioning that Kurucz' extended sequence of static stellar atmosphere models, thanks to its fine metallicity grid, has been linearly interpolated for providing the BC and the CTR at the quoted metallicities.

As a first step, let us compare these theoretical predictions with the observed C-M distribution of RR Lyrae stars in M3, as given by Sandage (1990). To allow such a comparison in Fig. 2 we present the $V$, $B - V$ topology of the instability strip by assuming a metal content $Z = 0.0004$, i.e. for a value of Z which should be suitable for this cluster. According to Bono, Caputo & Stellingwerf (1995), observed mean magnitudes and colors of variable stars have been corrected for the amplitudes to obtain the proper equilibrium values, i.e. the magnitude and color which a star would have if it were not pulsating. These values have been finally plotted in Fig. 2 assuming for the cluster a reddening $E(B - V) = 0$ and a distance modulus $DM = 14.95 \, mag$ (see Buonanno et al. 1994). On the basis of the quoted choice of $DM$ the observed lower envelope of the individual apparent magnitudes results in good agreement with the predicted absolute magnitude of the zero age horizontal branch within the instability strip ($M_V^{HB} = 0.76 \, mag$ at $Y = 0.24$ and $Z = 0.0004$). Moreover, the figure shows that the distribution of variables stars appears in good agreement with theoretical predictions. Indeed, the RR$ab$ variables populate the fundamental region and the region where both fundamental and first overtone are simultaneously excited ("*either-or*"), while the transition between RR*c* and RR*ab* takes place close to the FBE.

To perform a similar comparison for RR Lyrae stars belonging to M15 we will adopt observational data from Bingham et al. (1984). Fig. 3 shows the instability strip transformed into the observative plane by assuming the commonly adopted value of the cluster metallicity $Z = 0.0001$ (Zinn 1985). However, before plotting observational data it is also necessary to adopt a value for the M15 reddening. Although, this is a longstanding problem the quoted figure shows that by assuming $E(B-V) \sim 0.11$ as early suggested by Sandage, Katem & Sandage (1981) and recently supported by Walker (1994), no agreement can be found between theory and observations, since pulsating stars should invade the hot stable region of the diagram. On the contrary, by assuming $E(B-V) = 0.07$, as requested by Cacciari et al. (1984) and by adopting the canonical absolute magnitudes of 0.62 *mag* for HB stars in proximity of the HBE, the distribution of variables nicely overlaps the theoretical predictions. In this framework the RR*c* stars belonging to M15 populate both the first overtone and the "*either-or*" regions and the transition between RR*c* and RR*ab* takes place near the HRE, where almost all the double-mode variables appear to be concentrated.

In summary, a sound comparison between theoretical instability strips and observed CM distribution of RR Lyrae stars supports the suggestion that the region between the FBE and the HRE is populated by RR*c* stars in the case of M15 and by RR*ab* stars in the case of M3, confirming that in OoII clusters the RR*ab*/RR*c* transition occurs near the HRE, whereas in OoI clusters this transition takes place close to the FBE. This finding firmly supports the hypothesis that the evolutionary history of the variables, through a hysteresis mechanism in the pulsating mode, plays a key role in the understanding of Oosterhoff dichotomy (van Albada & Baker 1971; Caputo, Castellani & Tornambè 1989).

– 7 –

**Figure Captions**

Fig. 1.— Location in the HR diagram of the theoretical instability strip at fixed helium abundancy ($Y = 0.24$) and for two different mass values, $M = 0.75 M_\odot$ (solid line), and $M = 0.65 M_\odot$ (dashed line). From left to right, the lines represent the first overtone blue edge, the fundamental blue edge, the first overtone red edge, and the fundamental red edge.

Fig. 2.— Location in the Color-Magnitude diagram of the theoretical instability strips for $M = 0.65 M_\odot$ (solid line) and $M = 0.75 M_\odot$ (dashed line), transformed by assuming a metal content $Z = 0.0004$. The $RR_{ab}$ and $RR_c$ type variables belonging to M3 are shown as dots and circles respectively.

Fig. 3.— As in Fig. 2, but for a metal content $Z = 0.0001$ and RR Lyrae variables of M15. Solid line refers to $M = 0.75 M_\odot$ while dashed line refers to $M = 0.65 M_\odot$. Triangles represent double-mode pulsators, whereas nonvariable HB stars are shown as asterisks.

**Table 1.** *Fundamental and first overtone instability strip: theoretical plane.*

| $LOGL/L_\odot$ | $M = 0.65 M_\odot$ | | | | $M = 0.75 M_\odot$ | | | |
|---|---|---|---|---|---|---|---|---|
| | HBE (°K) | FBE (°K) | HRE (°K) | FRE (°K) | HBE (°K) | FBE (°K) | HRE (°K) | FRE (°K) |
| 1.91 | 6950 | 6950 | 6650 | 5750 | 7050 | 7050 | 6650 | 5750 |
| 1.81 | 7150 | 7050 | 6650 | 5850 | 7150 | 6950 | 6750 | 5850 |
| 1.72 | 7250 | 7050 | 6750 | 5950 | 7250 | 6950 | 6550 | 5950 |
| 1.61 | 7350 | 6950 | 6650 | 6050 | 7450 | 6650 | 6550 | 6050 |

**Table 2.** *Fundamental and first overtone instability strip: observative plane ($M_V$, $B-V$).*[a]

| | $M = 0.65 M_\odot$ | | | | | | | | | | | |
|---|---|---|---|---|---|---|---|---|---|---|---|---|
| | $Z = 10^{-4}$ | | | | $Z = 4 \times 10^{-4}$ | | | | $Z = 10^{-3}$ | | | |
| $M_V$ (mag) | HBE (mag) | FBE (mag) | HRE (mag) | FRE (mag) | HBE (mag) | FBE (mag) | HRE (mag) | FRE (mag) | HBE (mag) | FBE (mag) | HRE (mag) | FRE (mag) |
| .26 | .208 | .217 | .305 | .505 | .212 | .221 | .321 | .519 | .216 | .227 | .334 | .535 |
| .45 | .194 | .213 | .305 | .488 | .198 | .217 | .319 | .499 | .202 | .223 | .330 | .513 |
| .64 | .188 | .221 | .295 | .469 | .192 | .227 | .310 | .479 | .196 | .234 | .322 | .492 |
| .83 | .183 | .241 | .320 | .453 | .185 | .248 | .331 | .463 | .190 | .255 | .341 | .475 |
| | $M = 0.75 M_\odot$ | | | | | | | | | | | |
| .26 | .204 | .224 | .297 | .506 | .207 | .231 | .312 | .520 | .212 | .239 | .323 | .536 |
| .45 | .198 | .237 | .304 | .489 | .201 | .244 | .319 | .501 | .205 | .251 | .331 | .515 |
| .64 | .189 | .254 | .349 | .470 | .191 | .264 | .355 | .481 | .195 | .273 | .362 | .493 |
| .83 | .172 | .311 | .351 | .454 | .173 | .322 | .356 | .464 | .177 | .332 | .364 | .476 |

[a] Both the absolute visual magnitudes and the colors listed in this table are not directly referred to the luminosity levels and temperatures listed in Table 1. The magnitude levels were chosen in such a way as to cover the whole magnitude range, whereas the colors have been evaluated through a linear interpolation, at fixed magnitude level, of the color grid obtained by the temperature transformations.

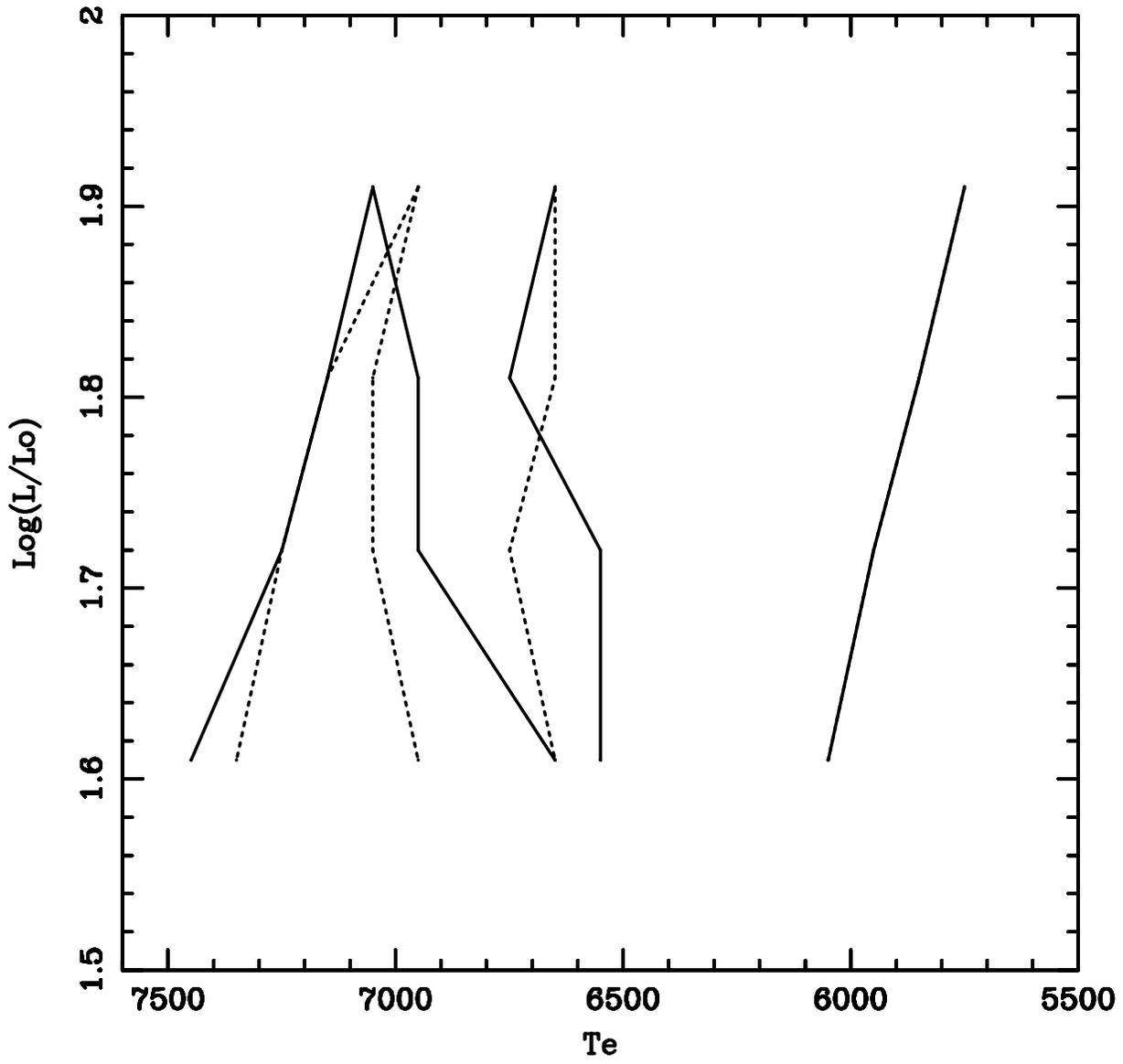

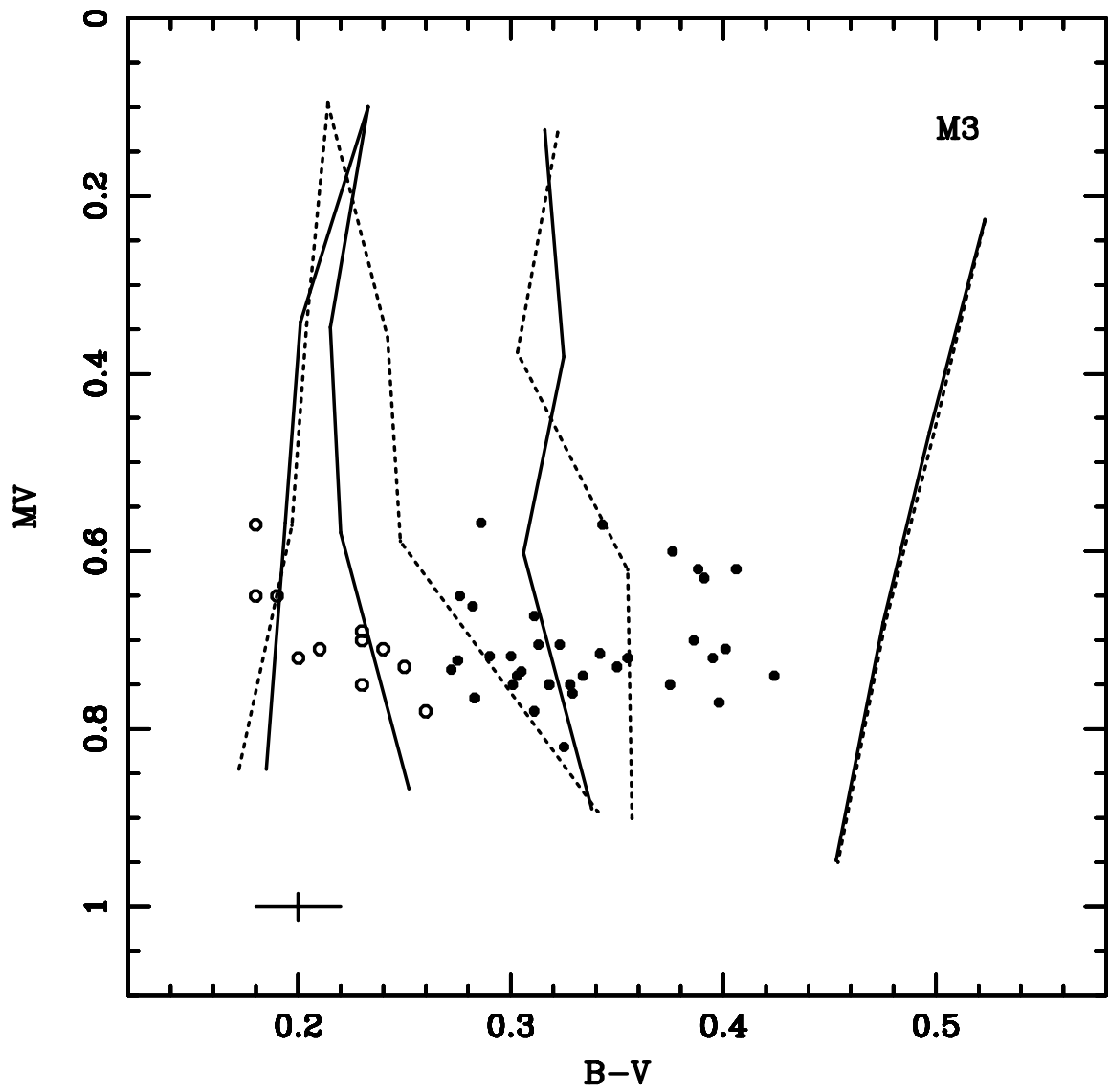

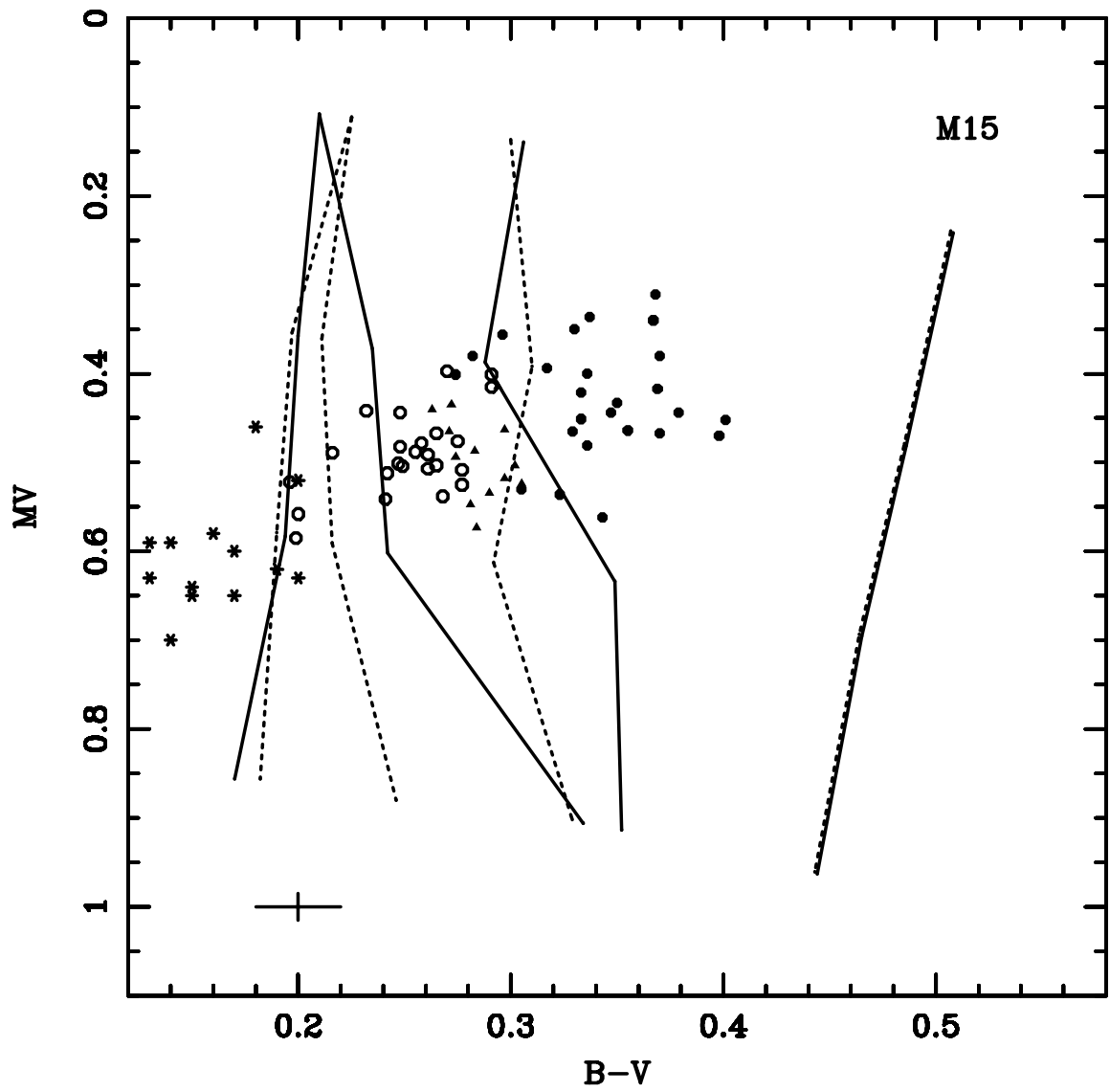